\documentclass[12pt,a4paper,draft]{article}

\usepackage{amstext,amsmath,amssymb,latexsym}

\def\nashi#1{}
\def\cl#1{{\cal#1}}
\def\ket#1{|#1\rangle}
\def\bra#1{\langle#1|}
\def\tr{{\rm Tr}}
\def\tran{{\,}^t}
\def\DCk{{D^{\Bbb C}_k}}

\newtheorem{proposition}{Proposition}
\newtheorem{lemma}{Lemma}
\newtheorem{theorem}{Theorem}

\begin{document}

\begin{center}
{\LARGE\bf
Bhattacharyya inequality for quantum state estimation}

\large
\bigskip
{Yoshiyuki \bf Tsuda}

\bigskip
COE, Chuo University \ \ \ \ \\
1-13-27 Kasuga, Bunkyo-ku, Tokyo 112-8551, Japan\bigskip

(The current address:
Institute of Statistical Mathematics\\
4-6-7 Minami-Azabu, Minato-ku, Tokyo 106-8569, Japan)
\end{center}

\begin{abstract}
Using higher-order derivative with respect to the
parameter, we will give lower bounds for variance of
unbiased estimators in quantum estimation problems. This
is a quantum version of the Bhattacharyya inequality in
the classical statistical estimation. Because of
non-commutativity of operator multiplication, we obtain
three different types of lower bounds; Type S, Type R and
Type L.
If the parameter is a real number,
the Type S bound is useful.
If the parameter is complex,
the Type R and L bounds are useful.
As an application, we will consider estimation of
polynomials of the complex amplitude of the quantum
Gaussian state. 
For the case where the amplitude lies in
the real axis, a uniformly optimum estimator for the
square of the amplitude will be derived using the Type S bound. 
It will be shown that there is no unbiased
estimator uniformly optimum as a polynomial of
annihilation and/or creation operators for the cube of the
amplitude. For the case where the amplitude does not
necessarily lie in the real axis, uniformly optimum
estimators for holomorphic, antiholomorphic and
real-valued polynomials of the amplitude will be derived.
Those estimators for the holomorphic and real-valued cases
attains the Type R bound, and those for the
antiholomorphic and real-valued cases attains the Type L
bound. 
This article clarifies what is the best method
to measure energy of laser.
\end{abstract}

\section{Introduction}

Quantum estimation is an important theory
in quantum information \cite{helstrom, holevo}.
It is not merely useful for many purposes,
for example, evaluation of realized
quantum information processing, 
but also is a fundamental problem in its own right.
In this theory,
we consider an optimization problem of measurements
estimating unknown state, with respect to
a risk function under appropriate restriction.
A typical case, we adopt in this article,
is minimization of variance
under unbiasedness condition.
What physicists call an observable
is an unbiased estimator,
and
the variance is equal to the mean square error
if the estimator is unbiased.

It has been known that
the quantum Cram\'er-Rao inequality gives a lower bound
based on first-order derivative and the Schwartz's
inequality
\cite{helstrom,holevo,yuen-lax}.
Although there is only one Cram\'er-Rao inequality
for classical statistical estimation \cite{lehmann},
there are several different inequalities
due to the non-commutativity of operator multiplication;
SLD-type, RLD-type and LLD-type inequalities
formulated by operators respectively called
Symmetric Logarithmic Derivative,
Right Logarithmic Derivative and
Left Logarithmic Derivative
\cite{holevo2001}.
Yuen and Lax \cite{yuen-lax} showed that,
for the quantum Gaussian state model,
the homodyne and heterodyne measurement, respectively,
uniformly attains
the SLD-type bound for the one-parameter model and
the RLD-type bound for the two-parameter model.
Nagaoka \cite{nagaoka} showed that the SLD bound
can be locally attained for all one-parameter models.
For asymptotic settings,
there are many arguments on the quantum Cram\'er-Rao bound
\cite{asymptotics,gill-massar}.

For the classical estimation problems,
Bhattacharyya \cite{bhattacharyya} has improved
the Cram\'er-Rao inequality
extending the order of the derivative.
For the classical Gaussian distribution model
with unknown mean parameter $\theta$,
the uniformly optimum estimator for any polynomial
$g(\theta)$
is made by the Hermite polynomials,
and it attains the classical Bhattacharyya bound.
See \cite{tanaka,tanaka-akahira}
for details in classical cases.
In the quantum estimation theory,
Brody and Hughston
\cite{brody1,brody2}
defined a Bhattacharyya-type
lower bound generalizing the SLD for pure states,
and they analyzed asymptotic property.

In this article,
first,
we will propose three quantum Bhattacharyya inequalities
for mixed quantum states with one real or complex
parameter.
Generalization of the SLD gives the Type S lower bound
for the real parameter case, and
generalization of RLD and LLD gives
the Type R and Type L bounds for the
complex parameter case.
Second,
as the application,
we will consider the quantum Gaussian state model
where the amplitude parameter $\theta$ is unknown.
If $\theta$ lies in the real axis,
the uniformly optimum estimator for $\theta^2$ attaining
the
Type S bound
is a self-adjoint observable given as a superposition of
the number counting operator and the homodyne operator.
This is realized by squeezing followed by the
number counting.
For $\theta^3$,
there is no uniformly optimum unbiased estimator
written as a polynomial of
the creation and/or annihilation operators.
If $\theta$ lies in the complex plain,
we will present
uniformly optimum operators for polynomials $g(\theta)$
of $\theta$ and $\bar\theta$ (the conjugate).
If $g(\theta)$ is holomorphic, i.e., $d g/d\bar\theta=0$,
then the optimum estimator is given by
the heterodyne measurement and it attains the Type R
bound.
If $g(\theta)$ is antiholomorphic, i.e., $d g/d\theta=0$,
then the optimum estimator is also given by
the heterodyne measurement and it attains the Type L
bound.
If $g(\theta)$ is real-valued, i.e.,
$g(\theta)=\overline{g(\theta)}$,
the optimum estimator is given by some polynomials
of the annihilation/creation operators
and it attains both Type R and Type L bounds.

In quantum optics,
the Gaussian state is a simple model of
laser whose complex amplitude is fluctuated
around $\theta\in\Bbb C$,
and
the energy of the laser is proportional
to $|\theta|^2$.
Using the Type R and L inequality,
we will see that the counting measurement is
optimum.
If $\theta\in\Bbb R$ is previously known,
the Type S inequality shows that
a counting measurement after a squeezing
operation is optimum.

This article is constructed as follows.
In Section 2, our problem will be formulated and
a known proposition of the quantum Cram\'er-Rao inequality
will be shown.
In Section 3,
new theoretical results of quantum Bhattacharyya
inequality
will be presented.
In Section 4,
our theory will be applied to the quantum Gaussian state
model.
Appendix A is the proof for Section 3,
and Appendix B is that for Section 4.

The author thanks the referees for useful comments.

\nashi
{
Talks on this topic have been presented
in Mathematical Society of Japan,
in Japanese Joint Statistical Meeting,
in a symposium at RIMS of Kyoto University,
in seminars at Tamagawa University,
University of Electro-Communications
and University of Tokyo.
The author thanks comments and encouragements
given there
by Professors
Masafumi Akahira,
Mitsuru Hamada,
Osamu Hirota,
Ken-ichi Koike,
Hiroshi Nagaoka,
Yoichi Nishiyama,
Tomohiro Ogawa,
Akimichi Takemura,
Fuyuhiko Tanaka,
Hidekazu Tanaka and
Nakahiro Yoshida.
The author also thanks Professors
Keiji Matsumoto and Masahito Hayashi
for useful discussion.
}

\section{Setup}

Suppose that there is a quantum system with an unknown state.
Consider a set of candidate states $\{\rho_\theta\}$
for the system parameterized by $\theta\in\Theta$.
In this article,
it is assumed that these density operators are invertible and
that $\Theta=\Bbb R$ or $\Theta=\Bbb C$.
We will use $\zeta$ as the parameter instead of $\theta$
when we need to remark that
the parameter may not be a real number.
Our interest is to estimate the true value of
$g(\theta)$, where $g:\Theta\to\Theta$ is a smooth function.
Since the probabilistic error is inevitable,
we consider an optimization problem of estimation under some
restriction.

An estimator $M$ for $g(\theta)$ is
a Positive Operator Valued Measure (POVM)
taking measurement outcomes in $\Theta$.
(In a strict definition,
an estimator should take measurement
outcomes in $g(\Theta)\supseteq\Theta$,
but, for theoretical convenience,
we adopt the weaker definition in this article.)

The expectation (=average=mean) of $M$ is
\[
	E[M]:=
	\int_{\omega\in\Theta}
	\omega \tr[\rho_\theta M(d\omega)]
	.
\]
If $E[M]=g(\theta)$ for any $\theta\in\Theta$,
$M$ is said to be unbiased.
We adopt the variance of $M$ as the risk function;
the variance is defined, in this article, as
\begin{align*}
	V[M]&:=
	\int_{\omega\in\Theta}
	|\omega-E[M]|^2
	\tr[\rho_\theta M(d\omega)]
	\\
	&=
	\int_{\omega\in\Theta}
	(\omega-E[M])\overline{(\omega-E[M])}
	\tr[\rho_\theta M(d\omega)]
	.
\end{align*}
An unbiased estimator with the minimum variance among all unbiased estimators
at a point $\theta\in\Theta$ is
said to be locally optimum at $\theta$.
If an unbiased estimator is optimum at any $\theta\in\Theta$,
it is said to be uniformly optimum.

The lower bound for the variance of unbiased estimators
has been
given by using the Schwartz's inequality
and the first-order derivative with respect to $\theta$.
This bound is called
the quantum Cram\'er-Rao inequality.
See
\cite{helstrom,holevo73,holevo73rao,holevo,holevo2001,yuen-lax}
for the proof and related topics.

\def\theproposition{}
\begin{proposition}
Assume that $\Theta=\Bbb R$,
the variance of any unbiased estimator $M$ for $g(\theta)$
satisfies
\begin{equation}
	\label{eq:qcr-s}
	V[M]\ge|g'(\theta)|^2/J^S
	.
\end{equation}
If $T:=g'(\theta)(J^S)^{-1}L^S+g(\theta)$
(A scaler $x$ is identified with $x$ times identity.)
is free of the parameter,
then the Projection Valued Measure (PVM)
taking measurement outcomes in $\Bbb R$
given by the self-adjoint
operator $T$ is the uniformly optimum unbiased estimator
for $g(\theta)$
and the equality holds for (\ref{eq:qcr-s}).

Assume that $\Theta=\Bbb C$,
the variance of any unbiased estimator $M$ for $g(\zeta)$
$(\zeta\in\Theta)$ satisfies
\begin{align}
	&
	V[M]\ge|g'(\zeta)|^2/J^R
	,
	\label{eq:qcr-r}
	\\
	&
	V[M]\ge|g'(\bar\zeta)|^2/J^L
	.
	\label{eq:qcr-l}
\end{align}
If $T:=g'(\zeta)(J^R)^{-1}L^R+g(\zeta)$
is free of the parameter
and if $T$ is normal,
i.e., $T T^\dag=T^\dag T$,
then the PVM
taking measurement outcomes in $\Bbb C$
given
by the spectrum decomposition of
$T$ is the uniformly optimum unbiased estimator
for $g(\zeta)$
and the equality holds for (\ref{eq:qcr-r}).
Similarly,
if $T:=\overline{g'(\bar\zeta)}(J^L)^{-1}L^L+g(\zeta)$ is free of the parameter and is normal,
it is the uniformly optimum unbiased estimator
and
the equality holds for (\ref{eq:qcr-l}).
\end{proposition}

Here, $J^S$, $L^S$, $J^R$, $L^R$, $J^L$ and $L^L$
are defined as follows.

\noindent
{\bf Definition of \boldmath $L^S$ and $J^S$.} \
For the case $\Theta=\Bbb R$,
let $L^S$ be a self-adjoint operator satisfying
\begin{equation}
	\frac{d}{d\theta}\rho_\theta=
	\frac{\rho_\theta L^S+L^S\rho_\theta}{2}
	,
	\label{eq:cr:s}
\end{equation}
and then define $J^S$ as $\tr[\rho_\theta (L^S)^2]$.
$L^S$ is called Symmetric Logarithmic Derivative (SLD)
and
$J^S$ is called SLD Fisher information.

\medskip
\noindent
{\bf Definition of \boldmath$L^R$, $L^L$, $J^R$ and $J^L$.} \
For the case $\Theta=\Bbb C$,
let $L^R$ and $L^L$ be operators satisfying
\begin{equation}
	\frac{d}{d\bar\zeta}\rho_\zeta
	=\rho_\zeta L^R
	,\qquad
	\frac{d}{d\zeta}\rho_\zeta
	=L^L \rho_\zeta
	\label{eq:cr:r}
\end{equation}
where $\zeta:=x+\sqrt{-1}y$,
$d/d\zeta:=(d/d x-\sqrt{-1}d/d y)/2$ and
$d/d\bar\zeta:=(d/d x+\sqrt{-1}d/d y)/2$
for real variables $x$ and $y$.
Then $J^R$ and $J^L$ are defined as
$\tr[\rho_\zeta L^R (L^R)^\dag]$ and
$\tr[L^L \rho_\zeta (L^L)^\dag]$.
$L^R$ is called Right Logarithmic Derivative (RLD)
and
$J^R$ is called RLD Fisher information.
Similarly,
$L^L$ is called Left Logarithmic Derivative (LLD)
and
$J^L$ is LLD Fisher information.

\bigskip
The definitions of $L^S$, $L^R$ and $L^L$
are not unique because
the equations (\ref{eq:cr:s}) and (\ref{eq:cr:r})
have many solutions on a space where
$\rho_\theta$ does not depend on $\theta$.
For example,
the heterodyne measurement for the Gaussian model
is obtained in the form
$(J^R)^{-1}L^R+\theta$ ($\theta\in\Bbb C$)
where $L^R$ is a solution in an extended system
with an ancilla state,
while the homodyne measurement
$(J^S)^{-1}L^S+\theta$ ($\theta\in\Bbb R$)
needs no extension.
In spite of the ambiguity of $L^\cdot$,
the inner product $J^\cdot$
is uniquely determined.

\bigskip
When the sample size is finite,
there is no unbiased estimators uniformly optimum
except for a few cases
where the parameter space is flat with respect to
the metric defined by $J^S$ \cite{matsumoto}.
When $\rho_\theta$ is not smooth with respect to $\theta$,
the difference instead of the derivative is useful
\cite{q-chapman}.

There may be cases where 
the unbiasedness condition is so strict that no estimator is unbiased, and where
the variance (or the means square error)
is not appropriate as the risk geometrically.
However, it is worth studying
such problems with a view to gaining 
theoretical insight.

\section{Quantum Bhattacharyya inequality}\label{sec:3}

The quantum Cram\'er-Rao inequality is generalized
by using higher-order derivative instead of
the first-order derivative.

\subsection{Quantum Bhattacharyya inequality of Type S for the real parameter case}
Consider the case $\Theta=\Bbb R$.
Let $L^S_k:=\tran(L_1,L_2,...,L_k)$ be a column vector of
self-adjoint operators satisfying
\begin{equation}
	\frac{d^k}{d\theta^k}\rho_\theta=
	\frac{\rho_\theta L_k+L_k\rho_\theta}{2}
	.
	\label{eq:k-sld}
\end{equation}

To simplify notations,
we introduce a column vector
$D_k:=\tran(d/d\theta,...,d^k/d\theta^k)$
of differential operators,
and we write (\ref{eq:k-sld}) as
\[
	D_k[\rho_\theta]=\frac{\rho_\theta L^S_k+L^S_k\rho_\theta}{2}
	.
\]
Let $J^S_k$ be a $k\times k$ matrix where the
$(i,j)$-th entry is
\[
	J_{i,j}:=\tr[\rho_\theta L_i L_j]
	.
\]
The definition of $J^S_k$ is also simplified as
$J^S_k=\tr_{k\times k}[\rho_\theta L^S_k \tran(L^S_k)]$
where $\tr_{m\times n}[A]$
means taking the trace of each entry
of an $m\times n$ matrix $A$, namely,
\[
	\tr_{m\times n}
	\left[
	\begin{pmatrix}
	A_{1,1} & \cdots & A_{1,n} \\
	\vdots  & \ddots & \vdots  \\
	A_{m,1} & \cdots & A_{m,n}
	\end{pmatrix}
	\right]
	=
	\begin{pmatrix}
	\tr[A_{1,1}] & \cdots & \tr[A_{1,n}] \\
	\vdots  & \ddots & \vdots  \\
	\tr[A_{m,1}] & \cdots & \tr[A_{m,n}]
	\end{pmatrix}
	.
\]
Though the definition of $L^S_k$ is not unique
for a system extension with a known ancilla state,
$J^S_k$ is uniquely determined.

Assume that $J^S_k$ is invertible.
The Quantum Bhattacharyya inequality of `Type S'
is given as follows.

\begin{theorem}
If $M$ is an unbiased estimator for $g(\theta)$,
it holds that
\begin{equation}
	V[M]\ge
	\tran(D_k[g(\theta)])
	(J^S_k)^{-1}
	D_k[g(\theta)]
	.
	\label{eq:th1}
\end{equation}
Especially,
if $T:=\tran D_k[g(\theta)](J^S_k)^{-1}L^S_k+g(\theta)$
is free of the parameter,
then the PVM $M$ given by
the self-adjoint observable $T$ is
the uniformly optimum unbiased estimator
and the equality holds for (\ref{eq:th1}).
\end{theorem}

See Appendix A for the proof.

\subsection{Quantum Bhattacharyya inequality of Type R and Type L for the complex parameter case}
Consider the case $\Theta=\Bbb C$.
Let $\DCk$ be a column vector
\[
	\DCk:=
	\tran\Big(
	\frac{d}{d\zeta},\frac{d}{d\bar\zeta},
	\frac{d^2}{d\zeta^2},
	\frac{d^2}{d\zeta d\bar\zeta},
	\frac{d^2}{d\bar\zeta^2},...,
	\frac{d^k}{d\zeta^k},...,
	\frac{d^k}{d\zeta^{k-l}d\bar\zeta^l},...,
	\frac{d^k}{d\bar\zeta^k}
	\Big)
	,
\]
where
\[
	\frac{d^m}{d\zeta^n d\bar\zeta^{m-n}}
	:=
	\frac{1}{2^m}
	\Big(
	\frac{d}{d x}-\sqrt{-1}\frac{d}{d y}
	\Big)^n
	\Big(
	\frac{d}{d x}+\sqrt{-1}\frac{d}{d y}
	\Big)^{m-n}
	.
\]
The number of the entries is
$K:=k(k+3)/2$.
Define column vectors $L^R_k$ and $L^L_k$ of $K$ operators
as the solutions to the equations
\[
	\DCk[\rho_\zeta]=\rho_\zeta L^R_k,\quad
	\DCk[\rho_\zeta]=L^L_k \rho_\zeta
	.
\]
%
Let $J^R_k$ and $J^L_k$ be $K\times K$ matrices given by
\[
	J^R_k:=
	\tr_{K\times K}
	[\rho_\zeta L^R_k (L^R_k)^\dag],\quad
	J^L_k:=\tr_{K\times K}[L^L_k \rho_\zeta (L^L_k)^\dag]
	,
\]
where
\[
	\begin{pmatrix}
	A_{1,1}&\cdots&A_{1,n}\cr
	\vdots&\ddots&\vdots\cr
	A_{m,1}&\cdots&A_{m,n}
	\end{pmatrix}^\dag
	:=
	\begin{pmatrix}
	A_{1,1}^\dag&\cdots&A_{m,1}^\dag\cr
	\vdots&\ddots&\vdots\cr
	A_{1,n}^\dag&\cdots&A_{m,n}^\dag
	\end{pmatrix}
	.
\]
Applying this notation to $\DCk$,
we define
\[
	\DCk^\dag:=
	\Big(
	\frac{d}{d\bar\zeta},\frac{d}{d\zeta},
	\frac{d^2}{d\bar\zeta^2},
	\frac{d^2}{d\zeta d\bar\zeta},
	\frac{d^2}{d\zeta^2},...,
	\frac{d^k}{d\bar\zeta^k},...,
	\frac{d^k}{d\zeta^l d\bar\zeta^{k-l}},...,
	\frac{d^k}{d\zeta^k}
	\Big)
	.
\]
The definitions of $L^R_k$ and $L^L_k$ are not unique
due to the system extension with a known ancilla state,
but those of $J^R_k$ $J^L_k$ are unique.

The Quantum Bhattacharyya inequalities
of Type R and Type L
are given as follows.
\begin{theorem}
If $M$ is an unbiased estimator for $g(\zeta)$,
then it holds that
\begin{align}
	V[M]&\ge
	\DCk^\dag[g(\zeta)]
	(J^R_k)^{-1}
	\DCk[g(\bar\zeta)]
	,
	\label{eq:th2:r}
	\\
	V[M]&\ge
	\DCk^\dag[g(\zeta)]
	(J^L_k)^{-1}
	\DCk[g(\bar\zeta)]
	.
	\label{eq:th2:l}
\end{align}
Especially,
if $T:=\DCk^\dag[g(\zeta)](J^R_k)^{-1}L^R_k+g(\zeta)$
is free of the parameter
and if
$T$ is normal,
then the PVM given by the spectrum decomposition of $T$ is
the uniformly optimum unbiased estimator
and the equality holds for (\ref{eq:th2:r}).
Similarly,
if $T:=\DCk^\dag[g(\zeta)](J^L_k)^{-1}L^L_k+g(\zeta)$ is free of the parameter,
it is the uniformly optimum unbiased estimator
and the equality holds for (\ref{eq:th2:l}).
\end{theorem}

See Appendix A for the proof.

\section{Application to the quantum Gaussian model}

For a known constant $N>0$
and an unknown parameter $\theta\in\Theta$,
let
\[
	\rho_\theta:=
	\frac{1}{\pi N}
	\int_{\alpha\in\Bbb C}
	\exp\Big(
	-\frac{|\alpha-\theta|^2}{N}
	\Big)
	\ket\alpha\bra\alpha
	d^2\alpha
	.
\]
Here,
$d^2\alpha$ means $d x d y$ where $\alpha=x+\sqrt{-1}y$,
and
$\ket\alpha$ is the coherent vector
of the complex amplitude $\alpha:=x+\sqrt{-1}y$, i.e.,
\[
	\ket\alpha:=
	\exp\Big(-\frac{|\alpha|^2}{2}\Big)
	\sum_{n=0}^\infty
	\frac{\alpha^n}{\sqrt{n!}}e_n
\]
where $\{e_n\}_{n=0}^\infty$ is the orthonormal system.

The quantum Gaussian model is a generalization of the
classical model of Gaussian distributions,
where the probability density is given as
\[
	f_\theta(x):=\frac{1}{\sqrt{2\pi}}
	\exp\Big(-\frac{(x-\theta)^2}{2}\Big)
	.
\]
Here, $\theta\in\Bbb R$ is unknown.
In the classical estimation problem of $\theta^k$,
the $k$-th Hermite polynomial
$T(x):=(-1)^k e^{x^2/2}(d^k/d x^k)e^{-x^2/2}$
is the uniformly optimum unbiased estimator which attains
the classical Bhattacharyya lower bound.

For the quantum Gaussian model,
we consider two models;
the real Gaussian model $\Theta=\Bbb R$
and
the complex Gaussian model $\Theta=\Bbb C$.
For the real Gaussian model,
we consider two cases;
$g(\theta)=\theta^2$ and $g(\theta)=\theta^3$.
For $g(\theta)=\theta^2$,
the optimum estimator is given by
a PVM with measurement outcomes in
$\Theta(=\Bbb R\supsetneq g(\Theta)=\{x\mid x\ge0\})$.
For $g(\theta)=\theta^3$,
it will be shown that any unbiased estimator
given by an observable as a polynomial of
the creation/annihilation operators can not
be uniformly optimum.
We will identify a self-adjoint operator with the PVM.

\begin{theorem}\label{th:3}
Suppose that $\Theta=\Bbb R$.

If $g(\theta)=\theta^2$, then the unbiased estimator
\begin{equation}
	\label{eq:th3}
	T=
	\frac{N(N+1)}{(2N+1)^2}(a^2+{a^\dag}^2)
	+
	\frac{N^2+(N+1)^2}{(2N+1)^2}a^\dag a
	-
	N\frac{N^2+(N+1)^2}{(2N+1)^2}
\end{equation}
uniformly attains the Type S lower bound,
so $T$ is uniformly optimum.
Here,
$a$ is the annihilation operator satisfying
$a e_n=\sqrt n e_{n-1}$ and $a a^\dag-a^dag a=I$ (identity).

If $g(\theta)=\theta^3$,
no unbiased estimator of the polynomial form
of the creation/annihilation operators
can be uniformly optimum.
\end{theorem}

See Appendix B.1 for the proof.

\bigskip
The value
$g(\theta)=\theta^2$ may be measured
by the counting measurement $a^\dag a-$constant,
or by the square of the homodyne measurement
$(a+a^\dag)^2-$constant.
This theorem says that the optimum estimator
is a superposition of these two measurements.
We also note that
this optimum estimator is realized as
the counting measurement
$b^\dag b-$constant
after the following squeezing operation
on the system;
\[
	\begin{pmatrix}a\\a^\dag\end{pmatrix}
	\mapsto
	\frac{1}{\sqrt{2N+1}}
	\begin{pmatrix}N+1&N\\N&N+1\end{pmatrix}
	\begin{pmatrix}a\\a^\dag\end{pmatrix}
	=:
	\begin{pmatrix}b\\b^\dag\end{pmatrix}
	.
\]

\bigskip
For the complex Gaussian model,
we will consider these three cases for a polynomial
$g(\zeta)$ of $\zeta\in\Bbb C$;
\smallskip\\
Holomorphic case: $d g(\zeta)/d\bar\zeta\equiv 0$,
\\
Antiholomorphic case: $d g(\zeta)/d\zeta\equiv 0$,
\\
Real-valued case: $g(\zeta)\equiv g(\bar\zeta)$.
\smallskip\\
In each case, an optimal unbiased estimator will be presented
by a PVM taking measurement outcomes in $\Bbb C$.

A PVM $M$ taking outcomes in $\Bbb C$ will be described by
a normal operator $T$, that is, $T T^\dag=T^\dag T$ and
\[
	T=\int_{\omega\in\Bbb C}\omega M(d\omega)
	.
\]
Since $E[M]=\tr[\rho T]$, it holds that
\[
	V[M]=
	\tr[\rho_\theta(T-E[T])(T-E[T])^\dag]=
	\tr[\rho_\theta(T-E[T])^\dag(T-E[T])]
	.
\]
See \cite{hayashi-sakaguchi} for details on normality and
subnormality of operators for quantum measurement.

For the holomorphic and antiholomorphic cases,
we need to extend the system to describe the
normal operators.
Let $\cl K$ be an ancilla system spanned by
$\{f_n\}_{n=0}^\infty$, and
let $b$ be the annihilation operator satisfying
$b f_n=\sqrt n f_{n-1}$ and $b b^\dag-b^\dag b=1$.
Namely,
the original annihilation operator $a$ means
$a\otimes I$,
and the new one $b$ may represent $I\otimes a$.
The original state $\rho_\zeta$ is extended to
$\rho_\zeta\otimes f_0 f_0^\dag$.

\begin{theorem}\label{th:4}
Suppose that $\Theta=\Bbb C$ and $g(\theta)$ is a polynomial of
$\theta$ and $\bar\theta$.

If $g(\theta)$ is holomorphic,
the unbiased estimator
\[
	T=g(a+b^\dag)
\]
uniformly attains the Type R lower bound,
so it is uniformly optimum.

If $g(\theta)$ is antiholomorphic,
the unbiased estimator
\[
	T=\bar g(a^\dag+b)
\]
uniformly attains the Type L lower bound
so it is uniformly optimum,
where $\bar g(z):=\overline{g(z)}$.

If $g(\theta)$ is real-valued,
the unbiased estimator
\begin{align*}
	T=&
	\sum_{m,n}
	c_{m,n}(N+1)^n
	\sum_{r=0}^{\min(m,n)}
	(-1)^{\min(m,n)-r}
	{\max(m,n)\choose\min(m,n)-r}
	\\
	&\times
	\frac{\min(m,n)!}{r!}
	\Big(\frac{a}{N+1}\Big)^{r+\max(0,n-m)}
	{a^\dag}^{r+\max(0,m-n)}
\end{align*}
uniformly attains both Type R and Type L lower bounds
so it is uniformly optimum,
where $g(z)=\sum_{m,n}c_{m,n}\theta^m\bar\theta^n$.
\end{theorem}

See Appendix B.2 for the proof.

\medskip
For the holomorphic and antiholomorphic cases,
the optimum estimators are realized
essentially by the heterodyne measurement,
that is,
the estimated values
for $g(a+b^\dag)$ and $\bar g(a^\dag+b)$ are
both obtained by
operating $g(\cdot)$ to the heterodyne outcome.
Hence they can be simultaneously carried out.
On the other hand,
for the real-valued case,
no ancilla system is used so that it can not be
measured simultaneously with the
holomorphic/antiholomorphic cases.

For a real-valued case $g(\theta)={\rm Re}(\theta)^2$,
the optimum estimator is of the form
$(a+a^\dag)^2/4-$constant,
i.e., the square of the homodyne measurement.
This measurement does not commute with that
for $g(\theta)=\theta^2$
$(\theta\in\Bbb R)$ of Theorem 3.

\

\

\noindent
{\LARGE\bf Appendices}

\bigskip
\noindent
The proofs for Theorems 1 and 2 will be given in
Appendix A,
and those for Theorems 3 and 4 will be given in
Appendix B.

\appendix
\section{Proofs of Theorems 1 and 2}

The first lemma implies that,
for any POVM estimator for $g(\theta)$,
there is a PVM which has the same expectation and
a smaller variance.

\begin{lemma}
Assume that $M$ is a POVM
taking measurement outcomes in $\Theta$.
Let
\[
	T=\int_{\omega\in \Theta}\omega M(d\omega)
	.
\]
Then, it holds that
\begin{align}
	\int_{\omega\in \Theta}
	|\omega|^2 \tr[\rho_\theta M(d\omega)]
	\ge
	\tr[\rho_\theta T T^\dag]
	,
	\label{eq:lem1}
	\\
	\int_{\omega\in \Theta}
	|\omega|^2 \tr[\rho_\theta M(d\omega)]
	\ge
	\tr[\rho_\theta T^\dag T]
	.
	\label{eq:lem2}
\end{align}
\end{lemma}
{\bf Proof.} \
The first formula (\ref{eq:lem1}) is obtained by
\begin{align*}
	&
	\int_{\omega\in \Theta}
	|\omega|^2 \tr[\rho_\theta M(d\omega)]
	-
	\tr[\rho_\theta T T^\dag]
	\\
	&
	=
	\tr
	\Big[
		\rho_\theta
		\int_{\omega\in \Theta}
		(\omega-T) (\bar\omega-T^\dag)
		M(d\omega)
	\Big]
	\ge0
	.
\end{align*}
Similarly, (\ref{eq:lem2}) is obtained by
\begin{align*}
	&
	\int_{\omega\in \Theta}
	|\omega|^2 \tr[\rho_\theta M(d\omega)]
	-
	\tr[\rho_\theta T^\dag T]
	\\
	&
	=
	\tr
	\Big[
		\rho_\theta
		\int_{\omega\in \Theta}
		(\bar\omega-T^\dag)(\omega-T)
		M(d\omega)
	\Big]
	\ge0,
\end{align*}
\hfill$\Box$

Therefore, for the proofs of Theorems 1 and 2,
it is sufficient to show that,
for the case $\Theta=\Bbb R$,
if $\tr[\rho_\theta T]=g(\theta)$ holds
for any $\theta\in\Theta$ then
\[
	\tr[\rho_\theta(T-g(\theta))^2]
	=
	\tr[\rho_\theta T^2]-g(\theta)^2
	\ge
	\tran D_k[g(\theta)]
	(J_k^S)^{-1}
	D_k[g(\theta)],
\]
and, for the case $\Theta=\Bbb C$,
if $\tr[\rho_\zeta T]=g(\zeta)$
and $\tr[\rho_\zeta T^\dag]=g(\bar\zeta)$ hold
for any $\zeta\in\Theta$ then
\begin{align*}
	&
	\tr[\rho_\zeta(T-g(\zeta))(T^\dag-g(\bar\zeta))]
	\\
	&\qquad=
	\tr[\rho_\zeta T T^\dag]-|g(\zeta)|^2
	\ge
	\DCk^\dag[g(\zeta)]
	(J_k^R)^{-1}
	\DCk[g(\bar\zeta)]
	,
	\\
	&
	\tr[\rho_\zeta(T^\dag-g(\bar\zeta))(T-g(\zeta))]
	\\
	&\qquad=
	\tr[\rho_\zeta T^\dag T]-|g(\zeta)|^2
	\ge
	\DCk^\dag[g(\zeta)]
	(J_k^L)^{-1}
	\DCk[g(\bar\zeta)]
	.
\end{align*}

If $\Theta=\Bbb R$, $T$ is a self-adjoint
operator, for which the existence of
the POVM is trivial.
On the other hand,
if $\Theta=\Bbb C$, one should consider
normality and/or subnormality of $T$ with
extension of the system.
See \cite{hayashi-sakaguchi}.

\subsection*{Proofs of Theorems 1 and 2.}\label{sec:a:1}

The Theorem 1 for $\Theta=\Bbb R$
and the Theorem 2 for $\Theta=\Bbb C$
are proved by using
the Schwartz's inequality in a similar way.

If, for any $\theta\in\Bbb R$,
an self-adjoint operator $T$ satisfies
$\tr[\rho_\theta T]=g(\theta)$, then
\begin{align*}
	&
	\tr_{k\times1}
	[\rho_\theta L^S_k(T-g(\theta))]
	=
	\tr_{k\times1}
	\Big[
	\frac{\rho_\theta L^S_k+L^S_k\rho_\theta}{2}
	(T-g(\theta))
	\Big]
	\\
	&=
	\tr_{k\times1}
	[D_k[\rho_\theta] T]
	-g(\theta)
	\tr_{k\times1}
	[D_k[\rho_\theta]
	=
	D_k[g(\theta)]
	.
\end{align*}
Let $U^S$ and $W^S$ be column vectors
of $k+1$ operators and $k+1$ scalers,
respectively,
given as
\[
	U^S:=
	\begin{pmatrix}
	T-g(\theta) \\
	L^S_k
	\end{pmatrix}
	,\qquad
	W^S:=
	\begin{pmatrix}
	1 \\
	-(J_k^S)^{-1}
	D_k[g(\theta)]
	\end{pmatrix}.
\]
Since
\[
	\Upsilon^S:=
	\tr_{k+1\,\times\,k+1}
	[\rho_\theta U^S (U^S)^\dag]
	=
	\begin{pmatrix}
	V[T] & \tran D_k[g(\theta)]\\
	D_k[g(\theta)] & J^S_k
	\end{pmatrix}
\]
is non-negative
where $V[T]:=\tr[\rho_\theta T^2]-g(\theta)^2$,
it holds that
\[
	\tran W^S (\Upsilon^S)^{-1} W^S
	=
	V[T]-\tran D_k[g(\theta)](J^S_k)^{-1}
	D_k[g(\theta)]\ge0
	.
\]
Hence we obtain the Theorem 1.

If, for any $\zeta\in\Bbb C$,
an operator $T$ satisfies
$\tr[\rho_\zeta T]=g(\zeta)$, then
\begin{align*}
\nashi
{
	&
	\tr_K
	[\rho_\zeta L^R_k(T^\dag-g(\bar\zeta))]
	=
	\tr_K
	[L^L_k \rho_\zeta(T^\dag-g(\bar\zeta))]
	\\
	&=
	\tr_K[\DCk[\rho_\zeta]T^\dag]
	-g(\bar\zeta)\tr_K[\DCk[\rho_\zeta]]
	=
	\DCk[g(\bar\zeta)],
	\\
}
	&
	\tr_K
	[\rho_\zeta(T-g(\zeta))(L^R_k)^\dag]
	=
	\tr_K
	[(T-g(\zeta))\rho_\zeta(L^L_k)^\dag]
	\\
	&=
	\tr_K[\DCk^\dag[\rho_\zeta]T]
	-g(\zeta)\tr_K[\DCk^\dag[\rho_\zeta]]
	=
	\DCk^\dag[g(\zeta)]
\end{align*}
where $K=k(k+3)/2$.
Let $U^R$ and $U^L$ be column vectors
of $K+1=(k+1)(k+2)/2$ operators,
and let
$W^R$ and $W^L$ be column vectors
of $K+1$ scalers,
given as
\begin{align*}
	&
	U^R:=
	\begin{pmatrix}
	T-g(\zeta) \\
	L^R_k
	\end{pmatrix}
	,\qquad
	U^L:=
	\begin{pmatrix}
	T-g(\zeta) \\
	L^L_k
	\end{pmatrix}
	,
	\\
	&
	W^R:=
	\begin{pmatrix}
	1 \\
	-(J_k^R)^{-1}
	\DCk[g(\bar\zeta)]
	\end{pmatrix}
	,\qquad
	W^L:=
	\begin{pmatrix}
	1 \\
	-(J_k^L)^{-1}
	\DCk[g(\bar\zeta)]
	\end{pmatrix}
	.
\end{align*}
Let
$V_1:=\tr[\rho_\zeta T T^\dag]-|g(\zeta)|^2$
and
$V_2:=\tr[\rho_\zeta T^\dag T]-|g(\zeta)|^2$.
Since
\begin{align*}
	\Upsilon^R:=&
	\tr_{K+1 \,\times\,K+1}
	[\rho_\zeta U^R (U^R)^\dag]
	=
	\begin{pmatrix}
	V_1[T] & \DCk^\dag[g(\zeta)]\\
	\DCk[g(\bar\zeta)] & J^R_k
	\end{pmatrix}
	,
	\\
	\Upsilon^L:=&
	\tr_{K+1 \,\times\,K+1}
	[U^L\rho_\zeta(U^L)^\dag]
	=
	\begin{pmatrix}
	V_2[T] & \DCk^\dag[g(\zeta)]\\
	\DCk[g(\bar\zeta)] & J^L_k
	\end{pmatrix}
\end{align*}
are non-negative,
it holds that
\begin{align*}
	(W^R)^\dag (\Upsilon^R)^{-1} W^R
	=&
	V_1[T]-\DCk^\dag[g(\zeta)](J^R_k)^{-1}
	\DCk[g(\bar\zeta)]\ge0
	\\
	\mbox{ and }
	(W^L)^\dag (\Upsilon^L)^{-1} W^L
	=&
	V_2[T]-\DCk^\dag[g(\zeta)](J^L_k)^{-1}
	\DCk[g(\bar\zeta)]\ge0
	.
\end{align*}
Hence
(\ref{eq:th2:r}) and (\ref{eq:th2:l})
of Theorem 2 are satisfied, respectively.
\hfill$\Box$

\section{Proofs of Theorems 3 and 4}

\begin{lemma}\label{lem:dif}
If
$P(\alpha):=\exp(-\alpha\bar\alpha/N)$
for $\alpha\in\Bbb C$,
then
\begin{align}
	&
	(\alpha-\zeta)^m(\bar\alpha-\bar\zeta)^n
	P(\alpha-\zeta)
	\ket\alpha\bra\alpha
	\nonumber
	\\
	&=
	\Big(\frac{N}{N+1}\Big)^m
	(a-\zeta)^n (a^\dag-\bar\zeta)^m
	P(\alpha-\zeta)\ket\alpha\bra\alpha
	\label{eq:gau:dif1}
	\\
	&=
	\Big(\frac{N}{N+1}\Big)^n
	\ket\alpha\bra\alpha
	(a-\zeta)^n (a^\dag-\bar\zeta)^m
	P(\alpha-\zeta)
	.
	\label{eq:gau:dif2}
\end{align}
\end{lemma}

\noindent
{\bf Proof.} \
Since $a=\sum_{i=0}^\infty\sqrt{i}e_{i-1}e_i^\dag$
and
$\ket\alpha=e^{-\alpha\bar\alpha/2}
\sum_i \alpha^i/\sqrt{i!}e_i$,
$a\ket\alpha=\alpha\ket\alpha$ and
$\bra\alpha a^\dag=\bar\alpha\bra\alpha$.
Moreover, since
\begin{align*}
	&
	\frac{d}{d\alpha}\ket\alpha
	=
	-\frac{\bar\alpha}{2}\ket\alpha
	+
	\exp\Big(-\frac{\alpha\bar\alpha}{2}\Big)
	\sum_{i=0}^\infty
	\frac{i}{\alpha}
	\frac{\alpha^i}{\sqrt{i!}}e_i
	\\
	&
	\frac{d}{d\bar\alpha}\bra\alpha
	=
	-\frac{\alpha}{2}\bra\alpha
	+
	\exp\Big(-\frac{\alpha\bar\alpha}{2}\Big)
	\sum_{i=0}^\infty
	\frac{i}{\bar\alpha}
	\frac{\bar\alpha^i}{\sqrt{i!}}e_i^\dag
	,
\end{align*}
we have
\begin{align}
	a^\dag \ket\alpha P(\alpha)
	=&
	\exp(-\alpha\bar\alpha/2)
	\sum_{i=0}^\infty
	\frac{i+1}{\alpha}
	\frac{\alpha^{i+1}}{\sqrt{(i+1)!}}
	e_{i+1} P(\alpha)
	\nonumber
	\\
	=&
	\ket\alpha
	\Big(\bar\alpha-\frac{d}{d\alpha}\Big)P(\alpha)
	,
	\label{eq:lem:gau:rule1}
	\\
	\bra\alpha a P(\alpha)
	=&
	\exp(-\alpha\bar\alpha/2)
	\sum_{i=0}^\infty
	\frac{i+1}{\bar\alpha}
	\frac{\bar\alpha^{i+1}}{\sqrt{(i+1)!}}
	e_{i+1}^\dag P(\alpha)
	\nonumber
	\\
	=&
	\ket\alpha
	\Big(\alpha-\frac{d}{d\bar\alpha}\Big)P(\alpha)
	\label{eq:lem:gau:rule2}
\end{align}
(for any smooth function $P(\alpha)$),
Recursively using these rules
(\ref{eq:lem:gau:rule1}) and (\ref{eq:lem:gau:rule2}) with
\[
	\frac{d}{d\zeta}
	P(\alpha-\zeta)
	=
	\frac{\bar\alpha-\bar\zeta}{N}
	P(\alpha-\zeta)
	\mbox{ and }
	\frac{d}{d\bar\zeta}
	P(\alpha-\zeta)
	=
	\frac{\alpha-\zeta}{N}
	P(\alpha-\zeta)
	.
\]
we obtain the results
(\ref{eq:gau:dif1}) and (\ref{eq:gau:dif2}).
\hfill$\Box$

\begin{lemma}\label{lem:lrll}
Let $p$ and $q$ be non-negative integers and
let $n:=\frac{(p+q-1)(p+q+2)}{2}+q+1$.
If $p\le q$,
the $n$-th entry of a Type R operator $L^R_k$ is
\begin{align}
	&
	\sum_{r=0}^{\min(p,q)}
	(-1)^{\min(p,q)-r}
	{\max(p,q)\choose \min(p,q)-r}\frac{\min(p,q)!}{r!}
	\nonumber
	\\
	&\times
	\frac
	{(a-\zeta)^{r+\max(0,q-p)}(a^\dag-\bar\zeta)^{r+\max(0,p-q)}}
	{N^p(N+1)^{r+\max(0,q-p)}}
	\label{eq:lem:lr}
\end{align}
The $n$-th entry of a Type L operator $L^L_k$ is
\begin{align}
	&
	\sum_{r=0}^{\min(p,q)}
	(-1)^{\min(p,q)-r}
	{\max(p,q)\choose \min(p,q)-r}\frac{\min(p,q)!}{r!}
	\nonumber
	\\
	&\times
	\frac
	{(a-\zeta)^{r+\max(0,q-p)}(a^\dag-\bar\zeta)^{r+\max(0,p-q)}}
	{N^q(N+1)^{r+\max(0,p-q)}}
	\label{eq:lem:ll}
\end{align}
\end{lemma}

\noindent
{\bf Proof.} \
If $p\le q$,
the higher order derivative
$(d^{p+q}/d^p\zeta d^q\bar\zeta)\rho_\zeta$ is calculated as
\begin{align}
	&
	\frac{1}{\pi N}
	\int_{\alpha\in\Bbb C}
	\ket\alpha\bra\alpha
	\frac{d^{p+q}}{d\zeta^p d\bar\zeta^q}
	\exp
	\Big(
	-\frac{(\alpha-\zeta)(\bar\alpha-\bar\zeta)}{N}
	\Big)
	d^2\alpha
	\nonumber
	\\
	=&
	\frac{1}{\pi N}
	\int_{\alpha\in\Bbb C}
	\ket\alpha\bra\alpha
	\frac{d^{p}}{d\zeta^p}
	\Big(\frac{\alpha-\zeta}{N}\Big)^q
	\exp
	\Big(
	-\frac{(\alpha-\zeta)(\bar\alpha-\bar\zeta)}{N}
	\Big)
	d^2\alpha
	\nonumber
	\\
	=&
	\frac{1}{\pi N}
	\int_{\alpha\in\Bbb C}
	\ket\alpha\bra\alpha
	\Big(\frac{d z}{d\zeta}\frac{d}{d z}\Big)^p
	\Big(\frac{z}{\bar\alpha-\bar\zeta}\Big)^q
	\exp\Big(-z\Big)
	d^2\alpha
	\nonumber
	\\
	&
	\big(z:=(\alpha-\zeta)(\bar\alpha-\bar\zeta)/N\big)
	\nonumber
	\\
	=&
	\frac{1}{\pi N}
	\int_{\alpha\in\Bbb C}
	\ket\alpha\bra\alpha
	\frac{(\bar\alpha-\bar\zeta)^{p-q}}{(-N)^p}
	\exp\Big(-\frac{|\alpha-\zeta|^2}{N}\Big)
	\nonumber
	\\
	&
	\times
	\sum_{r=0}^p
	(-1)^r
	{q\choose p-r}\frac{p!}{r!}
	\Big(\frac{|\alpha-\zeta|^2}{N}\Big)^{q-p+r}
	d^2\alpha
	.
	\label{eq:th3:long}
\end{align}
Similarly, if $p\ge q$,
\begin{align}
	\frac{d^{p+q}\rho_\zeta}{d^p\zeta d^q\bar\zeta}
	=&
	\frac{1}{\pi N}
	\int_{\alpha\in\Bbb C}
	\ket\alpha\bra\alpha
	\frac{(\alpha-\zeta)^{q-p}}{(-N)^q}
	\exp\Big(-\frac{|\alpha-\zeta|^2}{N}\Big)
	\nonumber
	\\
	&
	\times
	\sum_{r=0}^q
	(-1)^r
	{p\choose q-r}\frac{q!}{r!}
	\Big(\frac{|\alpha-\zeta|^2}{N}\Big)^{p-q+r}
	d^2\alpha
	.
	\label{eq:th3:long2}
\end{align}
By applying Lemma \ref{lem:dif} to
(\ref{eq:th3:long}) and (\ref{eq:th3:long2}),
the $n$-th entry of $L^R_k$ is obtained as
\[
	\begin{cases}
	\sum_{r=0}^p
	(-1)^{p-r}
	{q\choose p-r}\frac{p!}{r!}
	\frac
	{(a-\zeta)^{q-p+r}(a^\dag-\bar\zeta)^r}
	{N^p(N+1)^{q-p+r}}
	&
	\mbox{if }p\le q,
	\\
	\sum_{r=0}^q
	(-1)^{q-r}
	{p\choose q-r}\frac{q!}{r!}
	\frac
	{(a-\zeta)^r(a^\dag-\bar\zeta)^{p-q+r}}
	{N^p(N+1)^r}
	&
	\mbox{if }p\ge q,
	\end{cases}
\]
which means (\ref{eq:lem:lr}).
Likewise,
the $n$-th entry of $L^L_k$ is
\[
	\begin{cases}
	\sum_{r=0}^p
	(-1)^{p-r}
	{q\choose p-r}\frac{p!}{r!}
	\frac
	{(a-\zeta)^{q-p+r}(a^\dag-\bar\zeta)^r}
	{N^q(N+1)^r}
	&
	\mbox{if }p\le q,
	\\
	\sum_{r=0}^q
	(-1)^{q-r}
	{p\choose q-r}\frac{q!}{r!}
	\frac
	{(a-\zeta)^r(a^\dag-\bar\zeta)^{p-q+r}}
	{N^q(N+1)^{p-q+r}}
	&
	\mbox{if }p\ge q,
	\end{cases}
\]
which means (\ref{eq:lem:ll}).
\hfill$\Box$

\begin{lemma}\label{lem:char}
Suppose that $p,q,r,s$ are non-negative integers
and that
$p+q$ and $r+s$ are not larger than $k$.
Let
\[
	m:=\frac{(p+q-1)(p+q+2)}{2}+q+1
	,\quad
	n:=\frac{(r+s-1)(r+s+2)}{2}+r+1.
\]
Then the $(m,n)$-th entry of $J_k^R$ is
\begin{align}
	&
	\left.
	\frac
	{d^{p+q+r+s}}
	{d\kappa^p d\bar\kappa^q
	d\lambda^s d\bar\lambda^r}
	\exp
	\Big(
	\frac
	{\kappa\bar\lambda}
	{N}
	+
	\frac
	{\bar\kappa\lambda}
	{N+1}
	\Big)
	\right|_{\kappa=\lambda=0}
	\nonumber
	\\
	&=
	\begin{cases}
	\frac{p!q!}{N^p(N+1)^q}
	& \mbox{if }m=n,
	\\
	0 & \mbox{if }m\ne n,
	\end{cases}
	\label{eq:lem:mom}
\end{align}
and the $(m,n)$-th entry of $J_k^L$ is
\begin{align}
	&
	\left.
	\frac
	{d^{p+q+r+s}}
	{d\kappa^p d\bar\kappa^q
	d\lambda^s d\bar\lambda^r}
	\exp
	\Big(
	\frac
	{\kappa\bar\lambda}
	{N+1}
	+
	\frac
	{\bar\kappa\lambda}
	{N}
	\Big)
	\right|_{\kappa=\lambda=0}
	\nonumber
	\\
	=
	\begin{cases}
	\frac{p!q!}{N^q(N+1)^p}
	& \mbox{if }m=n,
	\\
	0 & \mbox{if }m\ne n.
	\end{cases}
	\label{eq:lem:mom:l}
\end{align}
\end{lemma}

\noindent
{\bf Proof.} \
Since $\rho_\zeta$ is invertible under the assumption $N>0$,
$L^R_k=\rho_\zeta^{-1}\DCk[\rho_\zeta]$ is a Type R operator
for any $\zeta\in\Bbb C$.
We have
\begin{align*}
	J^R_k
	=&
	\tr_{K\times K}[\rho_\zeta L^R_k(L^R_k)^\dag]
	\\
	=&
	\tr_{K\times K}
	\big[
	\rho_\zeta^{-1}\DCk[\rho_\zeta]\DCk^\dag[\rho_\zeta]
	\big]
\end{align*}
and hence the $(m,n)$-th entry is
\begin{equation}
	\label{eq:mom:tr}
	\left.
	\frac
	{d^{p+q+r+s}}
	{d\kappa^p d\bar\kappa^q
	d\lambda^s d\bar\lambda^r}
	\tr[\rho_\zeta^{-1}\rho_{\zeta+\kappa}\rho_{\zeta+\lambda}]
	\right|_{\kappa=\lambda=0}
	.
\end{equation}
The parameter $\zeta$ in (\ref{eq:mom:tr})
can be set to zero because
\[
	\tr[\rho_\zeta^{-1}\rho_{\zeta+\kappa}\rho_{\zeta+\lambda}]
	=
	\tr
	[U\rho_0^{-1}U^{-1}U\rho_\kappa U^{-1}U\rho_\lambda U^{-1}]
	=
	\tr[\rho_0^{-1}\rho_\kappa\rho_\lambda]
\]
where $U:=\exp(\zeta a^\dag-\bar\zeta a)$.
This can be calculated as
\begin{align}
	&
	\tr[
	\rho_0^{-1}\rho_\kappa\rho_\lambda]
	\nonumber
	\\
	=&
	\frac{N+1}{(\pi N)^2}
	\tr
	\Big[
	\sum_{n=0}^\infty
	\Big(\frac{N+1}{N}\Big)^n
	e_n e_n^\dag
	\cdot
	\int_{\alpha\in\Bbb C}
	\ket\alpha\bra\alpha
	\exp\Big(-\frac{|\alpha-\kappa|^2}{N}\Big)
	d^2\alpha
	\nonumber
	\\
	&
	\cdot
	\int_{\beta\in\Bbb C}
	\ket\beta\bra\beta
	\exp\Big(-\frac{|\beta-\lambda|^2}{N}\Big)
	d^2\beta
	\Big]
	\nonumber
	\\
	=&
	\frac{N+1}{(\pi N)^2}
	\int_{\alpha\in\Bbb C}
	\int_{\beta\in\Bbb C}
	\exp
	\Big(
	-\frac{|\alpha-\kappa|^2}{N}
	-\frac{|\beta-\lambda|^2}{N}
	+\frac{N+1}{N}\alpha\bar\beta
	\nonumber
	\\
	&
	-\alpha\bar\alpha-\beta\bar\beta+\bar\alpha\beta
	\Big)
	d^2\alpha d^2\beta
	\nonumber
	\\
	=&
	\frac{N+1}{(\pi N)^2}
	\int_{-\infty}^\infty
	\int_{-\infty}^\infty
	\int_{-\infty}^\infty
	\int_{-\infty}^\infty
	\exp
	\Big(
	-\tran(v-Q^{-1}\mu)Q(v-Q^{-1}\mu)
	\nonumber
	\\
	&
	+\frac{\kappa\bar\lambda}{N}
	+
	\frac{\bar\kappa\lambda}{N+1}
	\Big)
	d x d y d z d w
	\label{eq:int4}
\end{align}
where
$v:=\tran(x,y,z,w)$,
$\mu:=\tran\big({\rm Re}(\kappa),{\rm Im}(\kappa),
{\rm Re}(\lambda),{\rm Im}(\lambda)\big)/N$ and
\[
	Q:=
	\begin{pmatrix}
	1/N+1 & 0     & -1/(2N)-1 & -\sqrt{-1}/(2N) \\
	0     & 1/N+1 & \sqrt{-1}/(2N) & -1/(2N)-1 \\
	-1/(2N)-1 & \sqrt{-1}/(2N) & 1/N+1 & 0 \\
	-\sqrt{-1}/(2N) & -1/(2N)-1 & 0     & 1/N+1
	\end{pmatrix}
	.
\]
Since $Q>0$ and $\det Q=(N+1)^2/N^4$,
(\ref{eq:int4}) is equal to
\[
	\exp
	\Big(
	\frac{\kappa\bar\lambda}{N}+\frac{\bar\kappa\lambda}{N+1}
	\Big)
\]
so the result (\ref{eq:lem:mom}) is obtained.

Similarly,
$L^L_k$ can be given as
$L^L_k=\DCk[\rho_\zeta]\rho_\zeta^{-1}$.
Hence
\[
	J^L_k
	=
	\tr_{K\times K}[L^L_k\rho_\zeta(L^L_k)^\dag]
	=
	\tr_{K\times K}
	[\DCk[\rho_\zeta]\rho_\zeta^{-1}\DCk^\dag[\rho_\zeta]]
\]
so (\ref{eq:lem:mom:l}) holds.
\hfill$\Box$

\subsection{Proof of Theorem 3}

\subsubsection*{Optimality of (\ref{eq:th3}) for the case \boldmath$g(\theta)=\theta^2$}

First,
we formally extend the real parameter $\theta$ of $\rho_\theta$
to the complex parameter $\zeta=\theta+\sqrt{-1}\eta\in\Bbb C$.
Then the derivative
$(d/d\theta)^k\rho_\theta$ can be
considered as $(d/\zeta+d/d\bar\zeta)^k\rho_\zeta$.
A Type S operator $L^S_2=\tran(L_1,L_2)$ is given as a
solution to the equation
\begin{equation}
	\frac{d^k}{d\theta^k}\rho_\theta
	=
	\Big(\frac{d}{d\zeta}+\frac{d}{d\bar\zeta}\Big)^k\rho_\zeta
	=
	\frac{\rho_\theta L_k+L_k\rho_\theta}{2}
	\label{eq:th3:ls}
\end{equation}
and $L_k=L_k^\dag$ for $k=1,2$.
Let
\[
	L_k:=\sum_{i,j} c_{i,j}
	(a^i {a^\dag}^j+a^j {a^\dag}^i)
	\qquad
	(c_{i,j}\in\Bbb R)
	.
\]
Since each coefficient for $a^i {a^\dag}^j$ in
the equation (\ref{eq:th3:ls}) should be zero,
the solution is obtained as
\begin{align*}
	L_1
	=&
	\frac{2}{2N+1}(a+a^\dag-2\theta)
	\\
	=&
	2\frac{N+1}{2N+1}\frac{a-\theta}{N+1}
	+
	2\frac{N}{2N+1}\frac{a^\dag-\theta}{N}
	\\
	=&
	2\frac{N+1}{2N+1}M_{0,1}
	+
	2\frac{N}{2N+1}M_{1,0}
	,
	\\
	L_2
	=&
	\frac{2(a-\theta)^2+(a^\dag-\theta)^2}{N^2+(N+1)^2}
	+
	\frac{2(a-\theta)(a^\dag-\theta)}{N(N+1)}
	-
	\frac{2}{N}
	\\
	=&
	\frac{2(N+1)^2}{N^2+(N+1)^2}
	\Big(\frac{a-\theta}{N+1}\Big)^2
	+
	\frac{2N^2}{N^2+(N+1)^2}
	\Big(\frac{a^\dag-\theta}{N}\Big)^2
	\\
	&
	+
	2\Big(
	\frac{(a-\theta)(a^\dag-\theta)}{N(N+1)}
	-
	\frac{1}{N}
	\Big)
	\\
	=&
	\frac{2(N+1)^2}{N^2+(N+1)^2}
	M_{0,2}
	+
	\frac{2N^2}{N^2+(N+1)^2}
	M_{2,0}
	+
	2M_{1,1}
\end{align*}
where
$M_{i,j}$ are Type R operators out of
$L^R_2={}^t(M_{1,0},M_{0,1},M_{2,0},M_{1,1},M{0,2})$.
As the inner product $J^R_2$ of $L^R_2$ is given in
Lemma \ref{lem:char},
$J^S_2$ is obtained as
\[
	J^S_2=
	\begin{pmatrix}
	\frac{4}{2N+1} & 0 \\
	0 &
	\frac{8}{N^2+(N+1)^2}+\frac{4}{N(N+1)}
	\end{pmatrix}
	.
\]
Since $T$ of (\ref{eq:th3}) is equal to
$(2\theta,2)(J^S_2)^{-1}L^S_2+\theta^2$,
it attains the equality (\ref{eq:th1}),
hence it is uniformly optimum.
\hfill$\Box$

\subsubsection*{Non-existence of uniformly optimum estimator for the case \boldmath$g(\theta)=\theta^3$}

Since,
for all non-negative integers $m,n$,
the leading term of $\tr[\rho_\theta a^m (a^\dag)^n]$
is $\theta^m\bar\theta^n$,
the form of an unbiased estimator for $\theta^3$
of a polynomial form of the creation/annihilation operators
is
given in the form
\[
	T=
	u(a^3+{a^\dag}^3)+
	v({a^\dag}^2 a+a^\dag a^2)+
	w(a^2+{a^\dag}^2)+
	x a^\dag a +
	y (a+a^\dag)+
	z.
\]
Using the characteristic function
\begin{align*}
	\tr[\rho_\theta e^{\lambda a^\dag}e^{\bar\lambda a}]
	=&
	\frac{1}{\pi N}
	\int_{\alpha\in\Bbb C}
	\exp
	\Big(
	\lambda\bar\alpha+\bar\lambda\alpha
	-\frac{|\alpha-\theta|^2}{N}
	\Big)
	d^2\alpha
	\\
	=&
	\exp\big((\lambda+\bar\lambda)\theta+N|\lambda|^2\big)
	,
\end{align*}
we have
\[
	\tr[\rho_\theta T]=
	2 \theta^3 u+
	\big(2\theta^3+4\theta(N+1)\big) v+
	2 \theta^2 w+
	\big(\theta^2+N+1\big) x+
	2\theta y +
	z.
\]
The unbiasedness condition
$tbr[\rho_\theta T]=\theta^3$
requires that
\[
	u=\frac{1}{2}-v,\
	w=-\frac{x}{2},\
	y=-2(N+1)v,\
	z=-(N+1)x.
\]
It will be shown that,
for any fixed $\theta\in\Theta$,
the variance is minimized if
\begin{align}
	&
	u=
	\frac{N(N+1)}{2(4N^2+4N+3)},\quad
	v=
	3\frac{N^2+N+1}{2(4N^2+4N+3)},
	\nonumber
	\\
	&
	w=
	-\frac{3\theta}{2(2N+1)^2(4N^2+4N+3)},\quad
	x=
	\frac{3\theta}{(2N+1)^2(4N^2+4N+3)},
	\nonumber
	\\
	&
	y=
	-3\frac{(N^2+N+1)(N+1)}{4N^2+4N+3},\quad
	z=
	\frac{-3(N+1)\theta}{(2N+1)^2(4N^2+4N+3)}
	.
	\label{eq:uvwxyz}
\end{align}
Since $w$, $x$ and $z$ depend on $\theta$,
there is no unbiased estimator
uniformly minimizing the variance.

By the same way as the previous proof,
the third entry of
$L^S_3=\tran(L_1,L_2,L_3)$ can be obtained as
\begin{align*}
	L_3
	=&
	2
	\frac
	{(a-\theta)^3+(a^\dag-\theta)^3}
	{N^3+(N+1)^3}
	+
	6
	\frac
	{(a-\theta)^2(a^\dag-\theta)+(a-\theta)(a^\dag-\theta)^2}
	{N(N+1)(2N+1)}
	\\
	&
	-
	12
	\frac
	{a+a^\dag-2\theta}
	{N(2N+1)}
	\\
	=&
	\frac
	{2(N+1)^3}
	{N^3+(N+1)^3}
	\frac{(a-\theta)^3}{(N+1)^3}
	+
	\frac
	{2 N^3}
	{N^3+(N+1)^3}
	\frac{(a^\dag-\theta)^3}{N^3}
	\\
	&
	+
	\frac
	{6(N+1)^2N}
	{N(N+1)(2N+1)}
	\Big(
	\frac{(a-\theta)^2(a^\dag-\theta)}{(N+1)^2 N}
	-
	\frac{2(a-\theta)}{(N+1)N}
	\Big)
	\\
	&
	+
	\frac
	{6(N+1)N^2}
	{N(N+1)(2N+1)}
	\Big(
	\frac{(a-\theta)(a^\dag-\theta)^2}{(N+1)N^2}
	-
	\frac{2(a^\dag-\theta)}{N^2}
	\Big)
	\\
	=&
	\frac
	{2(N+1)^3}
	{N^3+(N+1)^3}
	M_{0,3}
	+
	\frac
	{2 N^3}
	{N^3+(N+1)^3}
	M_{3,0}
	\\
	&
	+
	\frac
	{6(N+1)^2N}
	{N(N+1)(2N+1)}
	M_{1,2}
	+
	\frac
	{6(N+1)N^2}
	{N(N+1)(2N+1)}
	M_{2,1}
\end{align*}
where $M_{i,j}$ are Type R operators out of
$L^R_3=\tran(M_{1,0},...,M_{3,0},M_{2,1},M_{1,2},M_{0,3})$.
Using Lemma \ref{lem:char},
we have
\[
	J^S_3=
	\begin{pmatrix}
	J^S_2 & \begin{array}{@{}c@{}}0\\0\end{array}\\
	\begin{array}{@{}cc@{}}0&0\end{array} &
	\frac{24}{N^3+(N+1)^3}
	+
	\frac{72}{N(N+1)(2N+1)}
	\end{pmatrix}
	.
\]
The unbiased estimator
satisfying formulas in (\ref{eq:uvwxyz})
is equal to
$T=\tran D_3[\theta^3](J^S_3)^{-1}L^S_3+\theta^3$,
which attains the lower bound (\ref{eq:th2:r})
for the variance at each $\theta$.
\hfill$\Box$

\subsection{Proof of Theorem 4}

\subsubsection*{The holomorphic case \boldmath$d g/d\bar\theta=0$}

Consider the monomial case $g(\theta)=\zeta^k$.
The column vector $\DCk[\bar\zeta^k]$ is given as
\[
	\DCk[\bar\zeta^k]=
	\tran
	\Big(0,k\bar\zeta^{k-1},0,0,k(k-1)\bar\zeta^{k-2},0,...,
	0,\frac{k!}{j!}\bar\zeta^j,0,...,
	0,k!\Big).
\]
Lemma \ref{lem:char} shows that
$J^R_k$ is a diagonal matrix of the form
\begin{align*}
	J^R_k
	=&
	{\rm diag}
	\Big(
	\frac{1}{N},\frac{1}{N+1},
	\frac{2}{N^2},\frac{1}{N(N+1)},
	\frac{2}{(N+1)^2},\frac{3!}{N^3},...,
	\\
	&
	\frac{(j-1)!}{N(N+1)^{j-1}},\frac{j!}{(N+1)^j},
	\frac{(j+1)!}{N^{j+1}},...,
	\frac{(k-1)!}{N(N+1)^{k-1}},\frac{k!}{(N+1)^k}
	\Big)
	.
\end{align*}
The system is extended with the ancilla space
$\cl K$, and the state is set as
$\rho_\zeta\otimes f_0f_0^\dag$.
The annihilation operator on $\cl K$ is
denoted by $b$.
A Type R operator $L^R_k$ on the extended system is
\begin{align*}
	L^R_k
	=&
	\tran\Big(
	\frac{a^\dag-\bar\zeta}{N},\frac{a+b^\dag-\zeta}{N+1},
	\frac{(a^\dag-\bar\zeta)^2}{N^2},
	\frac{(a+b^\dag-\zeta)(a^\dag-\bar\zeta)}{N(N+1)},
	\frac{(a+b^\dag-\zeta)^2}{(N+1)^2},
	\\
	&
	\frac{(a^\dag-\bar\zeta)^3}{N^3},...,
	\frac
	{(a+b^\dag-\zeta)^{j-1}(a^\dag-\bar\zeta)}{N(N+1)^{j-1}},
	\frac{(a+b^\dag-\zeta)^j}{(N+1)^j},
	\frac{(a^\dag-\bar\zeta)^{j+1}}{N^{j+1}},...,
	\\
	&
	\frac
	{(a+b^\dag-\zeta)^{k-1}(a^\dag-\bar\zeta)}{N(N+1)^{k-1}},
	\frac{(a+b^\dag-\zeta)^k}{(N+1)^k}
	\Big).
\end{align*}
Define an operator $T_k$ as
\[
	T_k
	:=
	\DCk^\dag[\zeta^k](J^R_k)^{-1}L^R_k+\zeta^k
	=
	\sum_{j=0}^k
	{k\choose j}
	\zeta^{k-j}(a+b^\dag-\zeta)^j
	=(a+b^\dag)^k
	.
\]
Since
$\tr_{K\times1}[\rho_\zeta L^R_k]=
\DCk[\tr[\rho_\zeta]]=\DCk[1]=0$,
$\tr[\rho_\zeta T_k]=\zeta^k$.
Hence $T_k$ is uniformly optimum unbiased estimator for
$\zeta^k$.
For the general holomorphic case
$g(\zeta)=\sum_k c_k \theta^k$ $(c_k\in\Bbb C)$,
the estimator $T:=g(a+b^\dag)=\sum_k c_k T_k$
is unbiased and uniformly optimum.
\hfill$\Box$

\subsubsection*{The antiholomorphic case \boldmath$d g/d\theta=0$}

Consider the monomial case $g(\theta)=\bar\zeta^k$.
The column vector $\DCk[\zeta^k]$ is given as
\[
	\DCk[\zeta^k]=
	\tran
	\Big(k\zeta^{k-1},0,k(k-1)\zeta^{k-2},0,...,
	0,\frac{k!}{j!}\zeta^j,0,...,
	0\Big).
\]
By Lemma \ref{lem:char},
$J^L_k$ is a diagonal matrix of the form
\begin{align*}
	J^L_k
	=&
	{\rm diag}
	\Big(
	\frac{1}{N+1},\frac{1}{N},
	\frac{2}{(N+1)^2},\frac{1}{N(N+1)},...,
	\\
	&
	\frac{(j-1)!}{N^{j-1}},
	\frac{j!}{(N+1)^j},
	\frac{(j-1)!}{N(N+1)^{j-1}},...,
	\frac{k!}{N^k}
	\Big)
	.
\end{align*}
For the annihilation operator $b$ on the ancilla system
for the extension $\rho_\zeta\otimes f_0 f_0^\dag$,
a Type L operator $L^L_k$ is
\begin{align*}
	L^L_k
	=&
	\tran\Big(
	\frac{a^\dag+b-\bar\zeta}{N+1},\frac{a-\zeta}{N},
	\frac{(a^\dag+b-\bar\zeta)^2}{(N+1)^2},
	\frac{(a-\zeta)(a^\dag+b-\bar\zeta)}{N(N+1)},...,
	\\
	&
	\frac{(a-\zeta)^{j-1}}{N^{j-1}},
	\frac{(a^\dag+b-\bar\zeta)^j}{(N+1)^j},
	\frac{(a-\zeta)(a^\dag+b-\bar\zeta)^{j-1}}{N(N+1)^{j-1}}
	,...,
	\\
	&
	\frac
	{(a-\zeta)^{k-1}(a^\dag+b-\bar\zeta)}{N^{k-1}(N+1)},
	\frac{(a-\zeta)^k}{N^k}
	\Big).
\end{align*}
Define an operator $T_k$ as
\[
	T_k
	:=
	\DCk^\dag[\bar\zeta^k](J^L_k)^{-1}L^L_k+\bar\zeta^k
	=
	\sum_{j=0}^k
	{k\choose j}
	\zeta^{k-j}(a^\dag+b-\bar\zeta)^j
	=(a^\dag+b)^k
	.
\]
Since
$\tr_{K\times1}[\rho_\zeta L^L_k]=
\DCk[\tr[\rho_\zeta]]=\DCk[1]=0$,
$\tr[\rho_\zeta T_k]=\bar\zeta^k$.
Hence $T_k$ is uniformly optimum unbiased estimator for
$\bar\zeta^k$.
For the general antiholomorphic case
$g(\zeta)=\sum_k c_k \bar\theta^k$ $(c_k\in\Bbb C)$,
the estimator $T:=g(a^\dag+b)=\sum_k c_k T_k$
is unbiased and uniformly optimum.
\hfill$\Box$

\subsubsection*{The real-valued case \boldmath$g=\overline g$}

By Lemmas \ref{lem:lrll} and \ref{lem:char},
$T_{m,n}:=
\DCk^\dag[c_{m,n}\zeta^m\bar\zeta^n](J^R_k)^{-1}
L^R_{m+n}+\zeta^m\bar\zeta^n$ for $k\ge m+n$
is
\begin{align*}
	T_{m,n}
	=&
	c_{m,n}
	\sum_{p=0}^n\sum_{q=0}^m
	\frac{m!n!\zeta^{m-q}\bar\zeta^{n-p}}{(m-q)!(n-p)!}
	\frac{N^p(N+1)^q}{p!q!}
	\\
	&\times
	\begin{cases}
	\displaystyle
	\sum_{r=0}^p
	(-1)^{p-r}
	{q\choose p-r}\frac{p!}{r!}
	\frac
	{(a-\zeta)^{q-p+r}(a^\dag-\bar\zeta)^r}
	{N^p(N+1)^{q-p+r}}
	& (p\le q)
	\\
	\displaystyle
	\sum_{r=0}^q
	(-1)^{q-r}
	{p\choose q-r}\frac{q!}{r!}
	\frac
	{(a-\zeta)^r(a^\dag-\bar\zeta)^{p-q+r}}
	{N^p(N+1)^r}
	& (p\ge q)
	\end{cases}
	\\
	=&
	c_{m,n}
	\sum_{p=0}^n\sum_{q=0}^m
	\frac{m!n!\zeta^{m-q}\bar\zeta^{n-p}}{(m-q)!(n-p)!}
	\\
	&\times
	\begin{cases}
	\displaystyle
	\sum_{r=0}^p
	\frac
	{(-1)^{p-r}(a-\zeta)^{q-p+r}(a^\dag-\bar\zeta)^r}
	{(p-r)!\,(q-p+r)!\,r!(N+1)^{-p+r}}
	& (p\le q)
	\\
	\displaystyle
	\sum_{r=0}^q
	\frac
	{(-1)^{q-r}(a-\zeta)^r(a^\dag-\bar\zeta)^{p-q+r}}
	{(q-r)!\,(p-q+r)!\,r!(N+1)^{-q+r}}
	& (p\ge q).
	\end{cases}
\end{align*}
$T_{m,n}$ does not depend on $\zeta\in\Bbb C$ because
$d T_{m,n}/d\zeta$ is equal to
\begin{align*}
	&
	c_{m,n}
	\sum_{p=0}^n\sum_{q=0}^{m-1}
	\frac{m!n!\zeta^{m-q-1}\bar\zeta^{n-p}}{(m-q-1)!(n-p)!}
	\\
	&\times
	\begin{cases}
	\displaystyle
	\sum_{r=0}^p
	\frac
	{(-1)^{p-r}(a-\zeta)^{q-p+r}(a^\dag-\bar\zeta)^r}
	{(p-r)!\,(q-p+r)!\,r!(N+1)^{-p+r}}
	& (p\le q)
	\\
	\displaystyle
	\sum_{r=0}^q
	\frac
	{(-1)^{q-r}(a-\zeta)^r(a^\dag-\bar\zeta)^{p-q+r}}
	{(q-r)!\,(p-q+r)!\,r!(N+1)^{-q+r}}
	& (p>q).
	\end{cases}
	\\
	&-
	c_{m,n}
	\sum_{p=0}^n\sum_{q=1}^m
	\frac{m!n!\zeta^{m-q}\bar\zeta^{n-p}}{(m-q)!(n-p)!}
	\\
	&\times
	\begin{cases}
	\displaystyle
	\sum_{r=0}^p
	\frac
	{(-1)^{p-r}(a-\zeta)^{q-p+r-1}(a^\dag-\bar\zeta)^r}
	{(p-r)!\,(q-p+r-1)!\,r!(N+1)^{-p+r}}
	& (p<q)
	\\
	\displaystyle
	\sum_{r=1}^q
	\frac
	{(-1)^{q-r}(a-\zeta)^{r-1}(a^\dag-\bar\zeta)^{p-q+r}}
	{(q-r)!\,(p-q+r)!\,(r-1)!(N+1)^{-q+r}}
	& (p\ge q)
	\end{cases}
	\\
	=&
	c_{m,n}
	\sum_{p=0}^n\sum_{q=0}^{m-1}
	\frac{m!n!\zeta^{m-q-1}\bar\zeta^{n-p}}{(m-q-1)!(n-p)!}
	\\
	&\times
	\begin{cases}
	\displaystyle
	\sum_{r=0}^p
	\frac
	{(-1)^{p-r}(a-\zeta)^{q-p+r}(a^\dag-\bar\zeta)^r}
	{(p-r)!\,(q-p+r)!\,r!(N+1)^{-p+r}}
	& (p\le q)
	\\
	\displaystyle
	\sum_{r=0}^q
	\frac
	{(-1)^{q-r}(a-\zeta)^r(a^\dag-\bar\zeta)^{p-q+r}}
	{(q-r)!\,(p-q+r)!\,r!(N+1)^{-q+r}}
	& (p>q).
	\end{cases}
	\\
	&-
	c_{m,n}
	\sum_{p=0}^n\sum_{q'=0}^{m-1}
	\frac{m!n!\zeta^{m-q'-1}\bar\zeta^{n-p}}{(m-q'-1)!(n-p)!}
	\\
	&\times
	\begin{cases}
	\displaystyle
	\sum_{r=0}^p
	\frac
	{(-1)^{p-r}(a-\zeta)^{q'-p+r}(a^\dag-\bar\zeta)^r}
	{(p-r)!\,(q'-p+r)!\,r!(N+1)^{-p+r}}
	& (p\le q')
	\\
	\displaystyle
	\sum_{r'=0}^{q'}
	\frac
	{(-1)^{q'-r'}(a-\zeta)^{r'}(a^\dag-\bar\zeta)^{p-q'+r'}}
	{(q'-r')!\,(p-q'+r')!\,r'!(N+1)^{-q'+r'}}
	& (p>q')
	\end{cases}
	\\
	=&
	0,
\end{align*}
where $q':=q-1$ and $r':=r-1$.
$T_{m,n}$ does not depend on $\bar\zeta$ because
$d T_{m,n}/d\bar\zeta$ is equal to
\begin{align*}
	&
	c_{m,n}
	\sum_{p=0}^{n-1}\sum_{q=0}^m
	\frac{m!n!\zeta^{m-q}\bar\zeta^{n-p-1}}{(m-q)!(n-p-1)!}
	\\
	&\times
	\begin{cases}
	\displaystyle
	\sum_{r=0}^p
	\frac
	{(-1)^{p-r}(a-\zeta)^{q-p+r}(a^\dag-\bar\zeta)^r}
	{(p-r)!\,(q-p+r)!\,r!(N+1)^{-p+r}}
	& (p<q)
	\\
	\displaystyle
	\sum_{r=0}^q
	\frac
	{(-1)^{q-r}(a-\zeta)^r(a^\dag-\bar\zeta)^{p-q+r}}
	{(q-r)!\,(p-q+r)!\,r!(N+1)^{-q+r}}
	& (p\ge q).
	\end{cases}
	\\
	&-
	c_{m,n}
	\sum_{p=1}^n\sum_{q=0}^m
	\frac{m!n!\zeta^{m-q}\bar\zeta^{n-p}}{(m-q)!(n-p)!}
	\\
	&\times
	\begin{cases}
	\displaystyle
	\sum_{r=1}^p
	\frac
	{(-1)^{p-r}(a-\zeta)^{q-p+r}(a^\dag-\bar\zeta)^{r-1}}
	{(p-r)!\,(q-p+r-1)!\,(r-1)!(N+1)^{-p+r}}
	& (p\le q)
	\\
	\displaystyle
	\sum_{r=0}^q
	\frac
	{(-1)^{q-r}(a-\zeta)^r(a^\dag-\bar\zeta)^{p-q+r-1}}
	{(q-r)!\,(p-q+r-1)!\,r!(N+1)^{-q+r}}
	& (p>q)
	\end{cases}
	\\
	=&
	c_{m,n}
	\sum_{p=0}^{n-1}\sum_{q=0}^m
	\frac{m!n!\zeta^{m-q}\bar\zeta^{n-p-1}}{(m-q)!(n-p-1)!}
	\\
	&\times
	\begin{cases}
	\displaystyle
	\sum_{r=0}^p
	\frac
	{(-1)^{p-r}(a-\zeta)^{q-p+r}(a^\dag-\bar\zeta)^r}
	{(p-r)!\,(q-p+r)!\,r!(N+1)^{-p+r}}
	& (p<q)
	\\
	\displaystyle
	\sum_{r=0}^q
	\frac
	{(-1)^{q-r}(a-\zeta)^r(a^\dag-\bar\zeta)^{p-q+r}}
	{(q-r)!\,(p-q+r)!\,r!(N+1)^{-q+r}}
	& (p\ge q).
	\end{cases}
	\\
	&-
	c_{m,n}
	\sum_{p'=0}^{n-1}\sum_{q=0}^m
	\frac{m!n!\zeta^{m-q}\bar\zeta^{n-p-1}}{(m-q)!(n-p-1)!}
	\\
	&\times
	\begin{cases}
	\displaystyle
	\sum_{r'=0}^{p'}
	\frac
	{(-1)^{p'-r'}(a-\zeta)^{q-p'+r'}(a^\dag-\bar\zeta)^{r'}}
	{(p'-r')!\,(q-p'+r')!\,r'!(N+1)^{-p'+r'}}
	& (p'<q)
	\\
	\displaystyle
	\sum_{r=0}^q
	\frac
	{(-1)^{q-r}(a-\zeta)^r(a^\dag-\bar\zeta)^{p'-q+r}}
	{(q-r)!\,(p'-q+r)!\,r!(N+1)^{-q+r}}
	& (p'\ge q)
	\end{cases}
	\\
	=&
	0,
\end{align*}
where $p':=p-1$ and $r':=r-1$.
Therefore,
\begin{align*}
	T_{m,n}
	=&
	c_{m,n}
	m!n!
	\begin{cases}
	\displaystyle
	\sum_{r=0}^p
	\frac
	{(-1)^{p-r}a^{q-p+r}(a^\dag)^r}
	{(p-r)!\,(q-p+r)!\,r!(N+1)^{-p+r}}
	& (p\le q)
	\\
	\displaystyle
	\sum_{r=0}^q
	\frac
	{(-1)^{q-r}a^r(a^\dag)^{p-q+r}}
	{(q-r)!\,(p-q+r)!\,r!(N+1)^{-q+r}}
	& (p\ge q)
	\end{cases}
	\\
	=&
	c_{m,n}(N+1)^n
	\sum_{r=0}^{\min(m,n)}
	(-1)^{\min(m,n)-r}
	{\max(m,n)\choose\min(m,n)-r}
	\\
	&\times
	\frac{\min(m,n)!}{r!}
	\Big(\frac{a}{N+1}\Big)^{r+\max(0,n-m)}
	{a^\dag}^{r+\max(0,m-n)}
	.
\end{align*}
Let $g(\zeta)=\sum_{m,n}$ be a real-valued polynomial,
namely, $c_{m,n}=\overline{c_{n,m}}$.
Let $T:=\sum_{m,n}c_{m,n}T_{m,n}$.
Then, $T$ is a self-adjoint observable.
Since $\tr_{k\times1}[\rho_\zeta L^R_k]=0$,
$\tr[\rho_\zeta T_{m,n}]=c_{m,n}\zeta^m\bar\zeta^n$.
Hence $T$ is an unbiased estimator for $g(\zeta)$.
$T$ attains the Type R lower bound because
$T=\DCk^\dag[g(\zeta)](J^R_k)^{-1}L^R_k$.

It can be shown in the same way that
$T=\DCk[g(\zeta)](J^L_k)^{-1}L^L_k$,
so that $T$ attains the Type L lower bound too.
\hfill$\Box$

\end{document}